\documentclass [12pt]{article}
\usepackage{epsfig}
\begin{document}

\begin{center}
\noindent{\Large\bf
Calculation of resonances
in the Coulomb three-body system with two disintegration channels
in the adiabatic hyperspherical approach
}\vspace{4mm}

\noindent{
{D.I. Abramov$^a$}, {V.V. Gusev$^b$}
}\vspace{1mm}

\noindent{
$^a$St. Petersburg State University\\
$^b$ Institute for High Energy Physics
}\vspace{4mm}
\end{center}

\bigskip

\noindent
{\bf Abstract.}
The method of calculation of the resonance characteristics
is developed for the metastable states
of the Coulomb three-body (CTB) system  with two disintegration channels.
It is based on the numerical solution of the scattering problem
in the framework of the adiabatic hyperspherical (AHS) approach.
The energy dependence of $K$-matrix in the resonance region
is calculated with the use of the stabilization method.
Resonance position $E_0$, partial widths $\Gamma_1$, $\Gamma_2$,
and three additional parameters
are obtained by fitting of the numerically calculated $K_{ij}(E)$
with the help of the generalized Breit-Wigner formula
which takes into account the non-zero background inelastic scattering.
The method developed  is applied to the calculation of the parameters
of three lowest metastable states
of the mesic molecular ion $dt\mu$.

\bigskip
\section*{Introduction}

The goal of the present paper is the extension
onto the two-channel case
of the numerical method developed previously
for the resonances in the CTB systems with one open channel
\cite{AG1}, \cite{AG2}.
In particular,
the special attention is devoted to the proper calculation
of the partial widths $\Gamma_i$
 which are interesting for applications
together with the total width
$\Gamma=\sum\Gamma_i$.

 The method developed is applied to three lowest
resonances in the $dt\mu$ system with zero angular momentum ($J=0$)
and two disintegration channels: (1) $(t\mu)_{n=1}+d$  and (2) $(d\mu)_{n=1}+t$.
They represent the quasistationary states in the potential well formed by
the third AHS term (fig.1).

\begin{figure}
\begin{center}
\epsfig{file=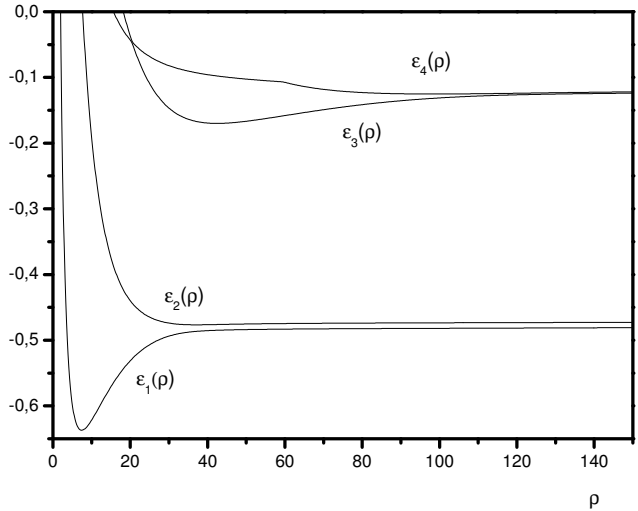,width=10cm}\end{center}

\caption{
AHS terms $\varepsilon_j(\rho)$ for the $dt\mu$ molecule. The resonances
under consideration are connected with the potential well in the third term.
}
\end{figure}

The resonances pointed out are interesting for many mesic atomic processes and
were investigated repeatedly.
The total widths $\Gamma=\Gamma_1+\Gamma_2$
were calculated in papers \cite{Hu}
\cite{Fro} (complex rotation method), \cite{Kin} (analysis of elastic
and inelastic scattering),
\cite{Tolst} (Siegert pseudo-states method), \cite{Mil} ($R$-matrix method).
The results of \cite{Hu}
\cite{Fro} differ significantly (3$\div$5 orders) from those of
\cite{Kin}, \cite{Tolst}, \cite{Mil}.
The results for the partial widths
which play an important role in the theory of the muon-catalyzed fusion
were presented up to now only in paper \cite{Kin} where
the non-adiabatic coupled rearrangement channel method
\cite{Kin2} had been used for the calculation of the elastic and inelastic cross sections.

Our method is based on the AHS
approach which is widely used in the calculations of mesic atomic systems
starting from paper \cite{G1}
(see for references
\cite{AGP}).
The three-body wave function is represented as a series over AHS basis, and
the determination of the reaction matrix $K_{ij}$ is reduced
to the numerical solution of
the system of radial equations (sec.1). It is calculated as a function
of the energy $E$ in the resonance region by the methods
described in paper \cite{AG1}. In particular, the stabilization method
\cite{Hazi} is used.

The resonance parameters
are obtained by the fitting numerical results for
$K_{ij}(E)$ with the help of the extended Breit-Wigner formula
presented in section 2. It takes into account
the nonzero background inelastic scattering and therefore
contains $(n+1)(n+2)/2$ independent parameters
($n$ is the number of open channels)
instead of $(2n+1)$ parameters in the well-known traditional
expression \cite{Landau}, \cite{Taylor}. Moreover,
this formula imposes the non-trivial restrictions on the values of $\Gamma_i$.
As it is seen from the results
for the lowest resonance in $dt\mu$ presented sec.4
the use of the extended formula
in this case
is necessary for an adequate representation
of $K_{ij}(E)$ and for a proper calculation of partial widths.

The method developed relates to ones
based on the numerical
solution of the scattering problem.
The typical feature of these methods
is the possibility to obtain the high accuracy for $\Gamma$
while  $E_0$ is calculated not so precisely \cite{AG1}.
Indeed, in the scattering problem
 $E_0$ and $\Gamma$
appear as the values of the entirely different nature:
$E_0$ is the position of the resonance peak while
$\Gamma^{-1}$ is its height, the errors of $\Gamma$
and $E_0$ are the independent values in a sense.
On the other hand
in the complex rotation method
the resonance  position $E_0$ and the half-width $\Gamma/2$
appear as the Cartesian
coordinates of the pole in the complex $E$-plane,
they are considered usually as the
values having the same rights and are calculated
with the same absolute error.

\section{Determination of $K$-matrix in AHSA}
We use the mesic atomic units
$
(\hbar=e=m_{\mu}=1)
$
and Jacobi coordinates
$$
{\bf R}={\bf r}_d-{\bf r}_t,\qquad
{\bf r}={\bf r}_{\mu}-\frac{m_{t}{\bf r}_{t}+m_d{\bf r}_d}
{m_{t}+m_d}.
\eqno(1)
$$
The hyperspherical
coordinates $\rho\in[0,\infty)$, $\chi\in[0\pi]$, $\vartheta\in[0,\pi]$
are defined by formulae
$$
  \rho^2 = 2\mu r^2 + 2MR^2,
$$
$$
  \tan\frac{\chi}{2} = \left(\frac{\mu}{M}\right)^{1/2}\frac{r}{R},\qquad
  \cos\vartheta = \frac{{\bf r R}}{rR} ,
$$
$$
  \mu^{-1} = 1+(m_{t}+m_{d})^{-1},\qquad
   M^{-1} = m_{t}^{-1}+m_{d}^{-1}.
   \eqno (2)
$$
The three-body Hamiltonian $H$ in the case $J=0$
 has the form \cite{G1}
$$
H= -{{1}\over{\rho^5}}{{\partial }\over{\partial \rho}}
\rho^5 {{\partial}\over{\partial \rho}} + h,
\eqno(3)
$$
where
the adiabatic Hamiltonian $h$
is given by the expression
$$
h=-
{{4}\over{\rho^2 \sin^2 \chi}}
\biggl(
{{\partial}\over{\partial \chi}} \sin^2 \chi
{{\partial}\over{\partial \chi}}
+\frac{1}{\sin{\vartheta}}\frac{\partial}{\partial\vartheta}
\sin{\vartheta}\frac{\partial}{\partial\vartheta}\biggr) +V(\rho,\chi,\vartheta),
\eqno(4)
$$
and $V$ is the sum of the Coulomb interactions between $t$, $d$ and $\mu$.

The three-body
wave function $~\Psi_E({\bf R},{\bf r})$
in the case $J=0$ depends only on $\rho$, $\chi$, $\vartheta$.
It is presented in the form of the AHS expansion
$$
\Psi_E({\bf R},{\bf r})=\rho^{-5/2}
         \sum_{j=1}^{\infty}
        f_j(\rho)\varphi_j(\rho|\chi,\vartheta),
\eqno (5)
$$
where the AHS basis functions
$\varphi_j(\rho|\chi,\vartheta)$, $j=1,2,...,\infty$
are the eigenfunctions of the AHS Hamiltonian $h$ (4):
$$
h\varphi_j(\rho|\chi,\vartheta)=\varepsilon_j(\rho)\varphi_j(\rho|\chi,\vartheta).
\eqno(6)
$$
Their normalization is defined by the relation
$$
\langle\varphi_i|\varphi_j\rangle \equiv
\int\limits_o^\pi \int\limits_o^\pi
\varphi_i(\rho|\chi,\vartheta)\varphi_j(\rho|\chi,\vartheta) ~\sin^2\chi
 \sin\vartheta d\chi d\vartheta=\delta_{ij}.
\eqno(7)
$$

Three lowest AHS energy terms $\varepsilon_j(\rho)$, $(j=1,2,3,4)$
are presented in fig.1, where their limiting values at $\rho\to\infty$, i.e.
the energies of bound states of corresponding atoms, are pointed out:
$$
\varepsilon_{1,3}(\infty)=E_{t\mu}(n=1,2),\qquad
\varepsilon_{2,4}(\infty)=E_{d\mu}(n=1,2).
\eqno(8)
$$
Radial functions
$f_j(E|\rho)$ satisfy
 the  infinite system of coupled ordinary differential equations
$$
  \left[-\frac{\partial^2}{\partial\rho^2}
  +\varepsilon_j(\rho)-E+\frac{15}{4\rho^2}\right]f_j +
$$
$$
  + \sum_{j'=1}^{\infty}\left[H_{jj'}f_{j'} +
     Q_{jj'}\frac{\partial}{\partial\rho}f_{j'} +
     \frac{\partial}{\partial\rho}\left(
     Q_{jj'}f_{j'}\right)\right] = 0,
     \eqno (9)
$$
$$
     H_{jj'} (\rho) =\langle\frac{\partial}{\partial\rho}\varphi_{j}|
    \frac{\partial}{\partial\rho}\varphi_{j'}\rangle,\qquad
    Q_{jj'} (\rho) =-\langle\varphi_{j}|
    \frac{\partial}{\partial\rho}\varphi_{j'}\rangle.
    \eqno (10)
$$

To calculate the two-channel reaction matrix
$K_{ij}(E)=K_{ji}(E)$, $(i,j=1,2)$
corresponding to channels
(1) $t\mu+d$ and (2) $d\mu+t$
with thresholds $E=\varepsilon_1(\infty)$ and $E=\varepsilon_2(\infty)$
one has to obtain two linearly independent solutions of system (9)
(we label them with upper index $i=1,2$)
with boundary conditions ($q_i^2=E-\varepsilon_i(\infty)$) \cite{AGP}:
$$
f^i_j(\rho){~\atop{\stackrel{=}{\rho \rightarrow 0}}}0,
~j=1,2,...,\infty;\qquad
f^i_j(\rho){~\atop{\stackrel{=}{\rho \rightarrow \infty}}}0,
~j=n+1,n+2,...,\infty;
$$
$$
f^i_j(\rho)
{~\atop{\stackrel{=}{\rho \rightarrow \infty}}}
\delta_{ij}\sin(q_j\rho-\pi J/2)+(q_i/q_j)^{1/2}K_{ij}\cos(q_j\rho-\pi J/2),
~j=1,2.
\eqno(11)
$$

The partial cross sections of elastic ($i=j$)
and inelastic ($i\neq j$) $s$-scattering ($J=0$)
is expressed in terms of $K_{ij}$ (left idex corresponds to input channel)\cite{AGP}:
$$
\sigma^{J=0}_{ij}=\frac {4\pi}{k^2_{i}}
\frac{\delta_{ij}D^2+K_{ij}^2}
{(1-D)^2+F^2},\qquad i,j=1,2,
\eqno(12)
$$
$$
D=K_{11}K_{22}-K_{12}K_{21},\qquad
F=K_{11}+K_{22},
$$
$$
 k_{i}=(2\mu_{i})^{1/2}q_{i} ,~~~\qquad
\mu_1^{-1} = (m_t +1)^{-1} + m_d^{-1} ,~~~\qquad
\mu_2^{-1} = (m_d +1)^{-1} + m_t^{-1} .
$$

\section{Generalized Breit-Wigner formula}

The general formula describing the $n$-channel scattering matrix
as a function of energy $E$ in the vicinity of
isolated complex pole $E_0-i\Gamma/2$ with small imaginary part
($\Gamma/ E_0\ll 1$)
can be derived from the unitary property, the symmetry and the
supposition that
 the residue of $S$-matrix in this pole
has an order $O(\Gamma)$.
In the special case
when the nonresonant inelastic scattering is absent and the background term
has a diagonal form
(we use the symbols with tilde for this special case)
this formula (the Breit-Wigner formula)
reads \cite{Landau}, \cite{Taylor}:
$$
\tilde S(E)=
\tilde S^b-\frac{i\Gamma \tilde B}{E-E_0+i\Gamma/2},
\eqno(13)
$$
where the background $S$-matrix $~\tilde S^b$ and the residue matrix
 $\tilde B$ do not depend on $E$ and are defined by
the expressions
$$
\tilde S^b_{ij}=\delta_{ij}e^{2i\Delta_i},\qquad
\tilde B_{ij}=\tilde \beta_i\tilde \beta_je^{i(\Delta_i+\Delta_j)}.
\eqno(14)
$$
Here
$\tilde \beta_i$ ($i=1,2,...,n$) are the real parameters (positive and negative)
 saisfying the condition
$$
\sum_{i=1}^n\tilde \beta_i^2=1.
\eqno(15)
$$

Thus, in the absence of the background inelastic scattering
the Breit-Wigner formula
contains $(2n+1)$ independent real parameters: $E_0$, $\Gamma$, independent
phaseshifts $\Delta_j$ ($j=1,2,...,n$), and parameters $\tilde\beta_j$ ($j=1,2,...,n$)
connected by relation (13).
The value
$$
\tilde\Gamma_j\equiv\Gamma\tilde \beta_j^2,\qquad j=1,...,n,\qquad
(\sum_{j=1}^n\tilde\Gamma_j=\Gamma)
\eqno(16)
$$
is the partial width corresponding to channel $j$.

In the general case, when
the background inelastic amplitude is not a negligibly small value,
it is necessary
to modify formula (13).
One has to find the general expression for
 matrix $B$ in the formula
$$
S(E)=S^b-\frac{i\Gamma B}{E-E_0+i\Gamma/2},
\eqno(17)
$$
where $S^b$ is
the symmetric unitary matrix which
may be the nondiagonal one.
This derivation can be easily done
with the use of
the orthogonal matrix $R$ reducing $S^b$ to the diagonal form:
$$
S^b=R^{-1}\tilde S^bR.
\eqno(18)
$$
Here the diagonal matrix $\tilde S^b$ is given by the first of eqs.(12),
so $\Delta_k$ ($k=1,...,n$) are the eigenphases of matrix $S^b$.
Now the orthogonal transformation  $S(E)=R^{-1}\tilde S(E)R$
reduces the problem of the search for the general form of $S(E)$
to the special case (13), and
as a result we obtain the general expression for $S(E)$ in the resonance range in the
form:
$$
S(E)=S^b-\frac{i\Gamma B}{E-E_0+i\Gamma/2}=
R^{-1}\left[\tilde S^b-\frac{i\Gamma \tilde B}{E-E_0+i\Gamma/2}\right]R,
\eqno(19)
$$
where $\tilde S^b$ and $\tilde B$ are given by eqs. (14),(15).

The general expression for $S(E)$ (19) thus contains
$(n+1)(n+2)/2$  real independent parameters: $E_0$, $\Gamma$,
independent phaseshifts $\Delta_j$, parameters
$\tilde\beta_j$  connected by eq.(15),
 and $n(n-1)/2$ parameters of an orthogonal  matrix $R$.

The matrix elements of $B$ in eqs.(17),(19) have the form
$$
B_{jk}=\sum_{l,m}R_{lj}\tilde B_{lm}R_{mk}=\sum_{l=1}^nR_{lj}\tilde \beta_le^{i\Delta_l}
\sum_{m=1}^nR_{mk}\tilde \beta_me^{i\Delta_m}=\beta_j\beta_ke^{i(\Delta_j+\Delta_k)},
\eqno(20)
$$
where $\beta_j$ $(~j=1,...,n)$ are the complex values
connected with parameters $\tilde \beta$ by the relation
$$
\beta_j=\sum_{l=1}^nR_{lj}\tilde \beta_le^{i(\Delta_l-\Delta_j)},
\qquad
(\sum_{i=1}^n|\beta_i|^2=1).
\eqno(21)
$$

The probability of the disintegration to the channel $i$ (partial width $\Gamma_i$)
is equal to
$$
\Gamma_i=\Gamma|\beta_i|^2,\qquad i=1,...,n,\qquad
(\sum_{i=1}^n\Gamma_i=\Gamma).
\eqno(22)
$$
It coincides with "eigenwidth" $\tilde\Gamma_i$
only if $R$ is an identity matrix.

The essential difference between the set $\{\Gamma_i\}$
and the set $\{\tilde\Gamma_i\}$ is that at fixed background scattering matrix
and complex pole
(i.e. at given $R$, $\Delta_j$, $E_0$ and $\Gamma$)
one can to define the positive parameters $\tilde\Gamma_1$ arbitrary
(with only restriction $\sum_{i=1}^n\tilde\Gamma_i=\Gamma$),
while the domain of definition of the parameters $\Gamma_i$ depends on $R$.
For example, in the case $n=2$ the transformation matrix has the form
$$
R=
\left(\begin{array}{cc}
\cos{\nu} &\sin{\nu} \\
&\\
-\sin{\nu} & \cos{\nu}\\
\end{array}\right),
\eqno(23)
$$
where $\nu$ is the mixing parameter.
The analysis of
the expression for $\beta_i$ (21) in this case leads to the following
restriction for $\Gamma_i/\Gamma=|\beta_i|^2$:
$$
2\Gamma_i/\Gamma\ge 1-\sqrt{\cos^2{(\Delta_1-\Delta_2)}
\sin^2{2\nu}+\cos^2{2\nu}}\ge 0.
\eqno(24)
$$
This inequality means that in the presence of the background inelastic scattering
($\nu\neq 0$) the probability of the disintegration certainly differs from zero
for every open channel.

The complex pole $(E_0-i\Gamma/2)$ of $S$-matrix corresponds to the real pole $E_1$
of the reaction matrix
$$
K=i(1+S)^{-1}(1-S)=K^b-
\frac{C}{E-E_1}.
\eqno(25)
$$
It is easy to obtain the expressions of $E_1$, background reaction
matrix $K^b$, and matrix $C$
in terms of parameters $E_0$, $\Gamma$, $\tilde \beta_j$, $\Delta_j$
and transformation matrix $R$:
$$
E_1=E_0-\frac{\Gamma}{2}\sum_{j=1}^n\tilde \beta_j^2\tan{\Delta_j},
\eqno(26)
$$
$$
K^b=R^{-1}\tilde K^bR,\qquad \tilde K^b_{jk}=\tan \Delta_j\delta_{jk},
\eqno(27)
$$
$$
C=R^{-1}\tilde CR,\qquad \tilde C_{jk}=
\frac{\Gamma}{2}\cdot\frac{\tilde \beta_j\tilde \beta_k}{\cos\Delta_j
\cos\Delta_k}.
\eqno(28)
$$

If we have to calculate
$E_0$, $\Gamma_i$, $\Gamma$
using the numerically obtained $ K(E)$
we have to fit $K(E)$ with the help of $(n+1)(n+2)/2$ real parameters
in accordance with eqs.(25)-(28).
In the case $n=2$ it is necessary to determine six such parameters.
We can use also
 six alternative parameters, namely, the pole of $K$-matrix $E_1$
 and the independent  matrix elements
 $K^b_{11}\equiv a_1$,  $K^b_{12}\equiv a$,
 $K^b_{22}\equiv a_2$, $C_{11}\equiv b_1$, $C_{22}\equiv b_2~$:
$$
K=
\left(\begin{array}{ll}
a_1 &a \\
&\\
a & a_2\\
\end{array}\right)-\frac{1}{(E-E_1)}
\left(\begin{array}{ll}
b_1 &b \\
&\\
b & b_2\\
\end{array}\right),
\eqno(29)
$$
where $b_1\ge 0$, $b_2\ge 0$, $b_1b_2-b^2=0$.

The case of the zero background inelastic scattering corresponds to $a=0$.

Resonance position $E_0$ and widths $\Gamma_i$, $\Gamma$ can be expressed
in the explicit form in terms of
$E_1$, $a_i$, $b_i$ with the help of eqs.(14)-(16), (26)-(28):
$$
E_0=E_1-\frac{h(1-d)-fg}{(1-d)^2+f^2},
\eqno(30)
$$
$$
\Gamma_1=2\frac{ha_2+b_1-db_2}{(1-d)^2+f^2},\qquad \Gamma_2=
2\frac{ha_1+b_2-db_1}{(1-d)^2+f^2},
\eqno(31)
$$
$$
\Gamma=\Gamma_1+\Gamma_2=2\frac{fh+g(1-d)}{(1-d)^2+f^2}.
\eqno(32)
$$
Here
$$
 d=a_{1}a_{2}-a^2,
\qquad f=a_{1}+a_{2},
$$
$$
  g=b_1+b_2,\qquad
h=a_1b_2+a_2b_1-2ab.
\eqno(33)
$$

\section{Numerical method}

Basis functions $\varphi_j(\rho|\chi,\vartheta)$ (6),(8) and matrix elements (10)
were calculated in accordance with algorithms described in papers \cite{G1}, \cite{ABGP}.
The calculations were performed on the
orthogonal finite-element grid $[N_\chi\times N_\vartheta]$ with
using the second order Lagrange elements.
The number of nodes in  $\chi$ and $\vartheta$  was taken equal to
$N_\chi=131$ and $N_\vartheta=61$.
This provided the accuracy  of calculation
$\sim 10^{-5}$ of all matrix elements.

The final results have been obtained with the number $N=6$ of
AHS basis functions.

The energy dependence of $K$-matrix (11) in the resonance range was obtained
with the use of stabilization method \cite{Hazi} in the same way as in papers \cite{AG1},
\cite{AG2}. We investigate the discrete spectrum of the auxiliary eigenvalue problem
for radial system (9) on the finite interval $0\le \rho\le \alpha$;
its eigenvalues $\Lambda_j(\alpha)$
have the avoided crossings at $\Lambda_j(\alpha)\approx E_0$ (see fig.2);
we solve the scattering problem for radial system (9) and
 calculate $K(E)$ at $E=\Lambda_j(\alpha)$ in the neighbourhood of the resonance;
 the scanning
along $\alpha$ instead of along $E$ allows to study in details the
behavior of $K(E)$ near $E=E_0$ (see fig.3).
\begin{figure}
\begin{center}
\includegraphics[angle=-90,width=10cm]{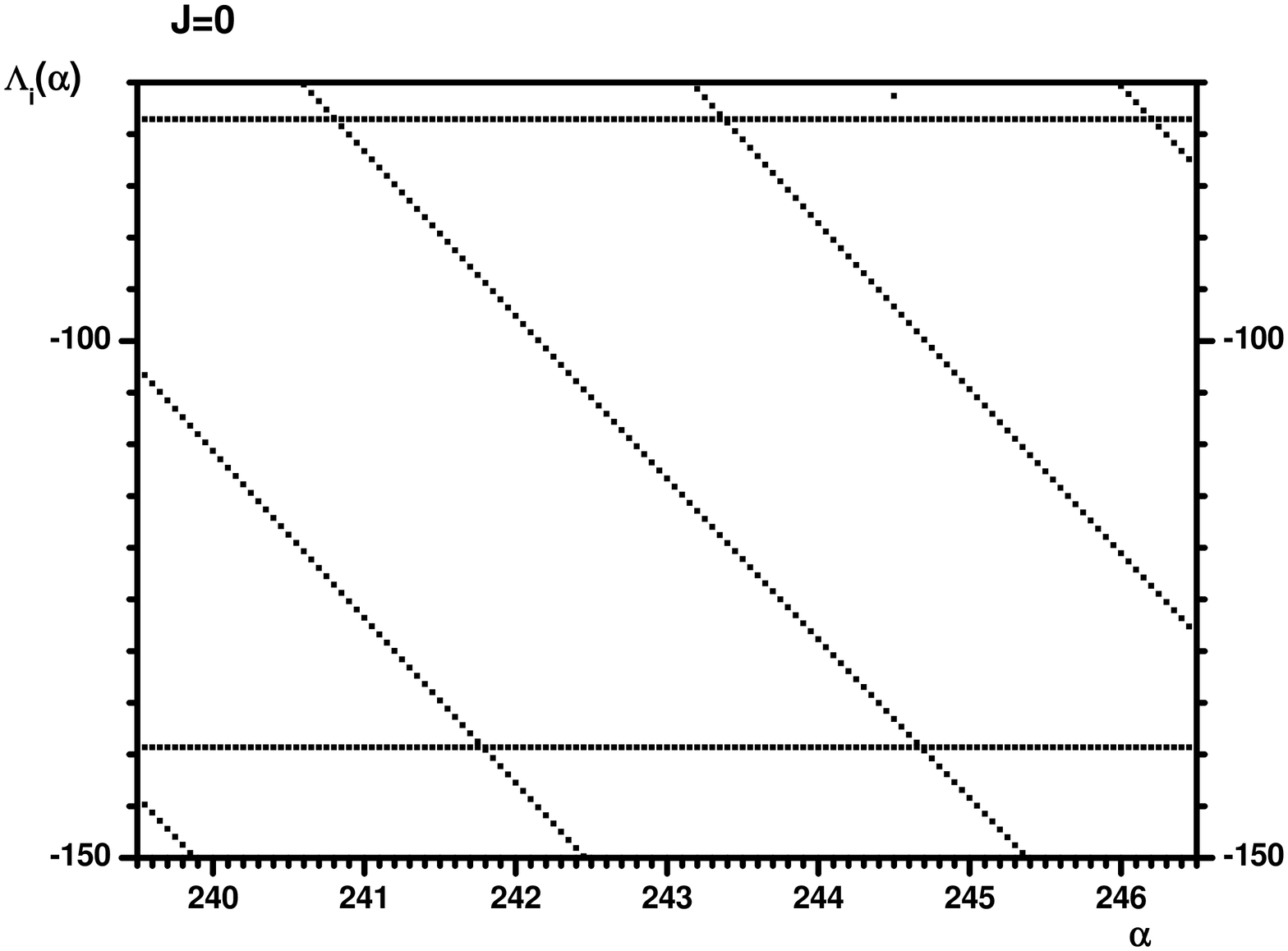}\end{center}

\caption{
Eigenvalues of the auxiliary boundary problem $\Lambda_j$ for $J=0$
as a function of right boundary $\alpha$.
The avoided crossings at $\Lambda_j=E_0$
appear as exact crossings on the scale of the figure.
}

\end{figure}

\begin{figure}
\begin{center}
\epsfig{file=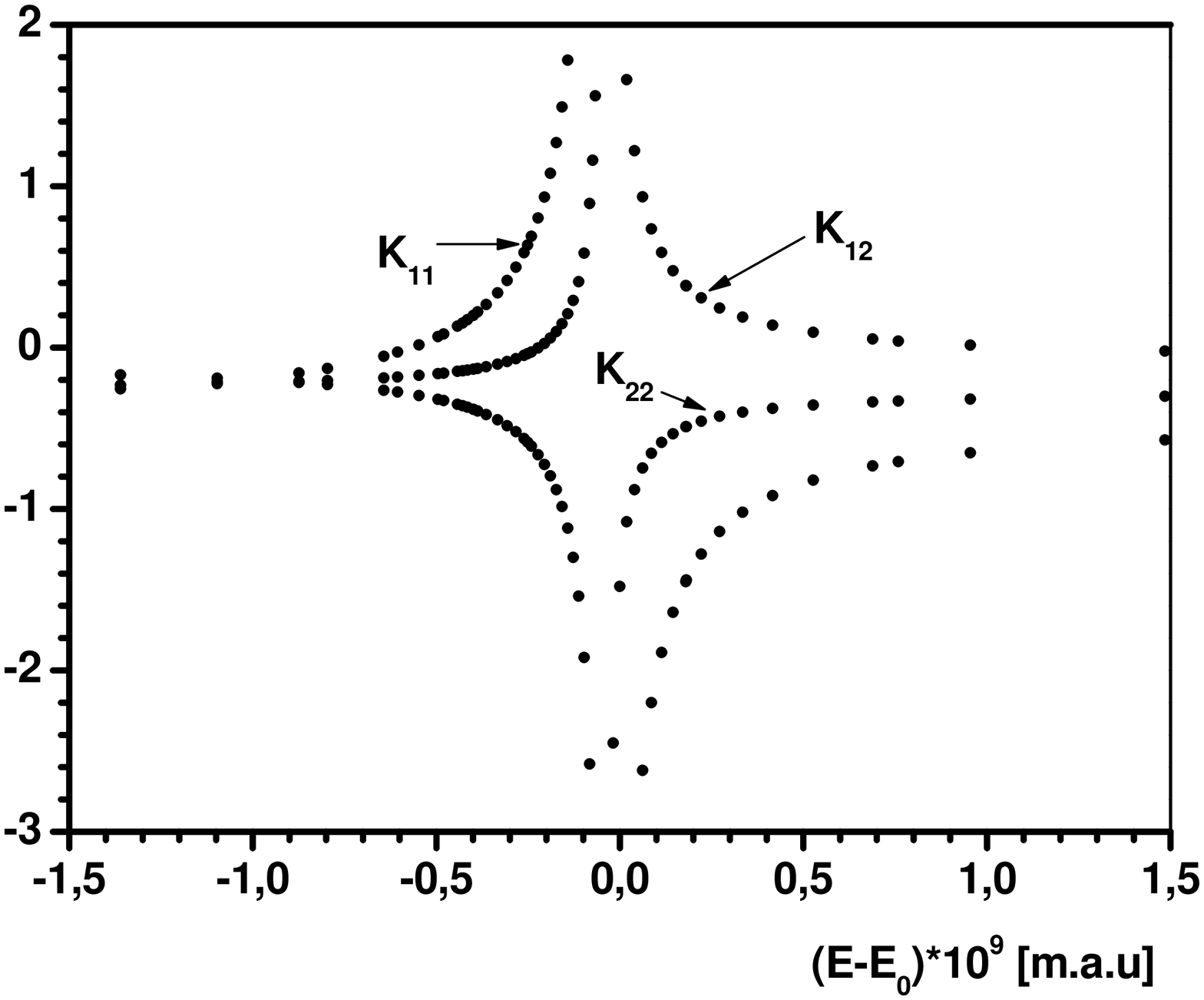,angle=-90,width=10cm}\end{center}

\caption{
 Matrix elements of $K$-matrix as functions of $E$
($J=0$, $v=0$).
}
\end{figure}

The system of radial equations (9) was solved at  $\rho\in[0.05,\rho^*=500$
with adapted step \cite{ABGP}.
This provides the results with the
relative accuracy $\sim 10^{-4}$.

Parameters $E_1,~a_1,~a_2,~a,~b_1,~b_2,~b$ of $K$-matrix (11),(29)
were obtained by fitting numerical results for $K(E)$ with the
help of the generalized Breit-Wigner formula (sec.2).
 The relation $b_1b_2b^{-2}=1$ was valid in this case with  accuracy $10^{-4}$.
Parameters $E_0$, $\Gamma$, $\Gamma_1$, and $\Gamma_2$ were calculated by
formulae (30)-(33).

As a whole we estimate the relative accuracy of calculation of $\Gamma$ and $\Gamma_i$
as $20\div 20$\%.

\section{Results and discussion}

\bigskip

The main results for three lowest resonances ($J=0$, $v=1,2,3$)
 are presented in figs. 3-5 and in tables 1, 2.

The numerically calculated matrix element  $K_{11}$ ($v=0$) as
function of $(E-E_0)$ is presented in fig. 3. Dots correspond to
the scanning with fixed step along $\alpha$ (sec.3). The step
along $E$ in this case
 becomes very small near resonance. The use of the scanning along $\alpha$
allows to trace the $E$-dependence
of $K_{11}$ in details.
\begin{figure}
\begin{center}
\epsfig{file=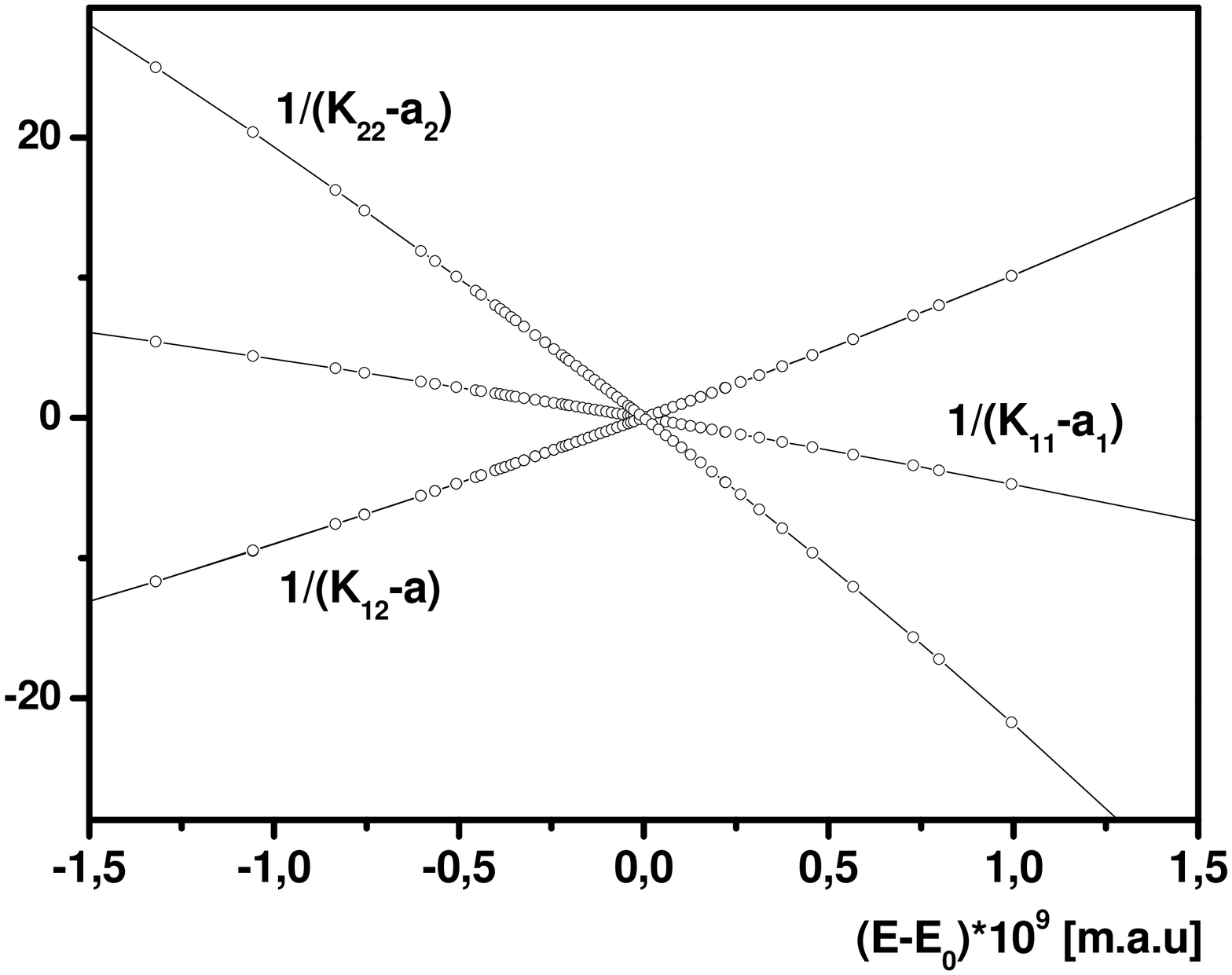,angle=-90,width=10cm}\end{center}

\caption{
 The inverse values of matrix elements of the resonant part of
 $K$-matrix ($J=0$, $v=0$).
}
\end{figure}

Fig.4 presents the numerical data for $(K_{11}-a_1)^{-1}$, $(K_{12}-a)^{-1}$, and
$(K_{22}-a_2)^{-1}$ as functions of $(E-E_0)$ for $v=0$.
It is seen, in particular, that all matrix elements of
$K$-matrix have the pole in the same point $E_0$.

The numerically obtained
elastic $\sigma_{11}$, $\sigma_{22}$ and inelastic $\sigma_{12}$
 cross sections (12) are presented on fig.5 as functions of the energy.
All curves have a typical  Fano form. The parameters of these
three profiles can be expressed in terms of $E_1$, $a_1$, $a_2$,
$a$, $b_1$, and $b_2$. The numerical data for $\sigma_{12}$ show
clearly that it can not be presented without taking into account
the nonzero background scattering. Indeed, at $a=0$ the profile
$\sigma_{12}(E)$ should be the symmetric one with respect to
$E=E_0$.

The data presented in tables 1 show that for three lowest resonances in the
$dt\mu$ system the use of $N=6$ AHS basis functions
gives for $E_0$ the accuracy 0.01\%. To improve the accuracy and to perform
the calculations for large $J$ and $v$ it is necessary
to increase $N$ and $\rho^*$.

Our results for $\Gamma$ (table 2) coincide with those of papers
\cite{Tolst}, \cite{Mil} with an accuracy 10$\div$50\%.
Independently of the precision of the numerical calculation of
reaction matrix $K(E)$ the results for branching ratio can change
essentially (about 10\% in our case) if one uses for data
processing the Breit-Wigner formula without taking into account an
inelastic background scattering, i.e. if one assumes $a=0$ in
eqs.(3)-(33).
\begin{figure}
\begin{center}
\epsfig{file=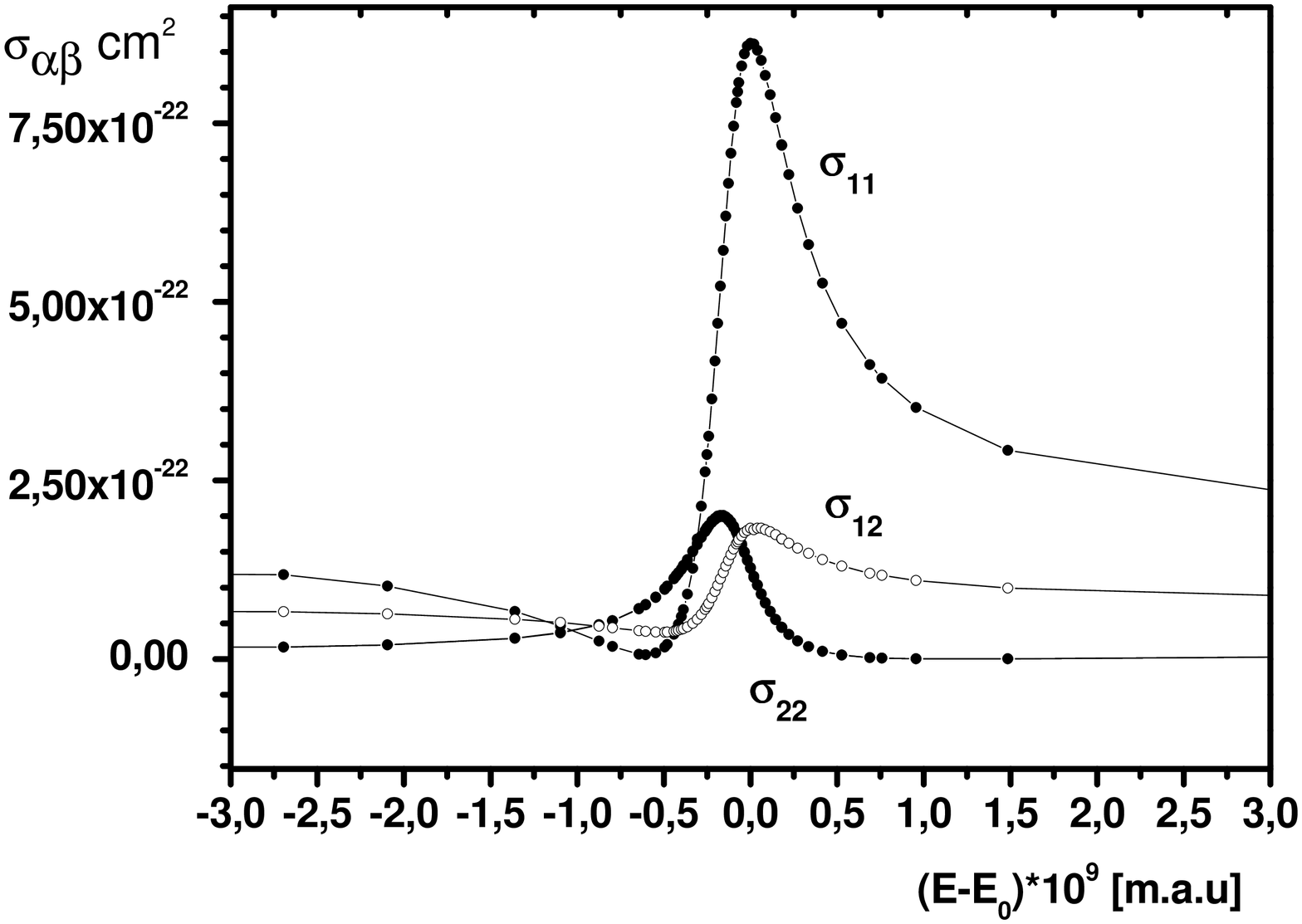,angle=-90,width=10cm}\end{center}

\caption{
Elastic and inelastic cross sections
in the resonance range ($J=0$, $v=0$).
}
\end{figure}

\bigskip

\section*{Acknowledgements}

This work was supported by the grant "Universities of Russia"
UR.01.01.064.

\newpage

\bigskip
\begin{center}

{\bf Table 1.}
Resonance position $-E_0$
relative to $n=2$ level of $t\mu$ (in $\mu$a.u.)\\
 for lowest resonances ($J=0$, $v=0,1,2$) in $dt\mu$ .

\vspace{5mm}

\begin{tabular}{|c|c|c|c|c|}

\hline\hline

$v$&\cite{Kin}&\cite{Tolst}&\cite{Mil}&Present\\
&&N=40&N=30$\div$50&N=6\\

\hline

0~~~~&0.159194&0.1591938&0.1591939&0.15917\\
1~~~~&0.145302&0.1453015&0.1453019&0.14524\\
2~~~~&0.134530&0.1345291&0.134529&0.13445\\

\hline\hline
\end{tabular}
\end{center}

\vspace{2cm}

\begin{center}

{\bf Table 2.}Total width $\Gamma$ ($10^{-9}\mu$a.u.)
and the ratio $\Gamma_2/\Gamma$ (in parenthesis)\\
 for lowest resonances ($J=0$, $v=0,1,2$) in $dt\mu$

\vspace{5mm}

\begin{tabular}{|c|c|c|c|c|c|c|}

\hline\hline

$v$&\cite{Hu}&\cite{Fro}&\cite{Kin}&\cite{Tolst}&\cite{Mil}&Present\\
&&&&N=40&N=30$\div$50& N=6\\
\hline 0&0.64$\times 10^4$&0.36$\times 10^3$&0.640 (0.1)&0.354
&0.341 &0.47 (0.22)\\
1&0.14$\times 10^5$&0.50$\times 10^4$&5.56~ (0.9)&0.839 &0.829 &0.53 (0.38)\\
2&0.20$\times 10^6$&0.12$\times 10^5$&37~~~ (0.9)&1.15~ &1.14~ &1.02 (0.55)\\

\hline\hline
\end{tabular}
\end{center}

\end{document}